\documentclass{ajour}

\usepackage{epsf}
\usepackage{cite}

\begin{document}

%


\authorrunninghead{Pascal Cedraschi and Markus B\"{u}ttiker}

\titlerunninghead{Quantum Coherence of the Ground State}





\title{Quantum Coherence of the Ground State of a Mesoscopic Ring}

\author{Pascal Cedraschi and Markus B\"{u}ttiker}

\affil{D\'epartement de Physique Th\'eorique,
Universit\'e de Gen\`eve,\\
24, quai Ernest Ansermet, CH-1211 Geneva 4, Switzerland}

\email{buttiker@serifos.unige.ch}

\abstract{We discuss the phase coherence properties of a mesoscopic
normal ring coupled to an electric environment via Coulomb
interactions.  This system can be mapped onto the Caldeira-Leggett
model with a flux dependent tunneling amplitude.  We show that
depending on the strength of the coupling between the ring and the
environment the free energy can be obtained either by a Bethe ansatz
approach (for weak coupling) or by using a perturbative expression
which we derive here (in the case of strong coupling).  We demonstrate
that the zero-point fluctuations of the environment can strongly
suppress the persistent current of the ring below its value in the
absence of the environment.  This is an indication that the
equilibrium fluctuations in the environment disturb the coherence of
the wave functions in the ring.  Moreover the influence of quantum
fluctuations can induce symmetry breaking seen in the polarization of
the ring and in the flux induced capacitance.}


\begin{article}

\section{Introduction}
\begin{figure}
\centerline{\epsfysize=5cm\epsfbox{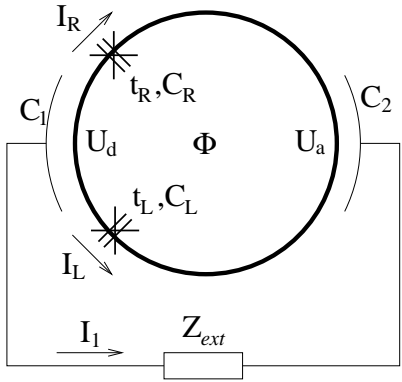}}
\vspace*{0.3cm}
\caption{\label{system}Ring with an in-line dot subject to a flux
$\Phi$ and capacitively coupled to an external impedance $Z_{ext}$.}
\end{figure}

Mesoscopic physics is concerned with systems that are so small that
the wave nature of the electrons becomes manifest.  The quantum
mechanical coherence properties of small electrical systems are thus a
central topic of this field.  In this work, we are interested in the
coherence properties of the ground state of a mesoscopic system,
namely a ring subject to an Aharonov-Bohm flux, which is capacitively
coupled to an external environment which exhibits zero-point
polarization fluctuations \cite{cedraschi:prl}.  In a mesoscopic ring,
quantum coherent electron motion leads, in the presence of an
Aharonov-Bohm flux, to a ground state that supports a persistent
current \cite{buettiker:AB,levy:manyrings,chandrasekhar:singlering,%
mailly:singlering,deblock:screening}.  The persistent current is thus
a signature of the coherence properties of the ground state of a
mesoscopic ring.  Only electrons whose wave function reaches around
the ring contribute to the persistent current.  Thermal fluctuations
are known to reduce the amplitude of the persistent current
\cite{landauer85,buettiker:squid,buttiker85}.  On the other hand, the
coherence properties of the ground state have thus far not been
discussed.  It is interesting to ask to what extent zero-point
fluctuations of an electrical environment can influence the ground
state of a ring, and how they affect the persistent current.  This
question is also interesting in view of the recent discussion on
whether or not weak localization properties are influenced by
zero-point fluctuations \cite{mohanty:decoh,golubev:decoh,%
aleiner:decoh:comment,golubev:decoh:reply,aleiner:decoh}.  This
discussion focuses on the conductance, a transport coefficient, and
not directly on the ground state of the system.  In this paper, we
extend and clarify the results presented in \cite{cedraschi:prl}.  A
more intuitive discussion of the results of \cite{cedraschi:prl}
based on a Langevin approach is given in \cite{cedraschi:langevin}.
Here we provide a detailed derivation based on Bethe ansatz results
and perturbation theory of the results obtained in
\cite{cedraschi:prl}.  In addition, we investigate the capacitance
coefficients of the ring and the charge-charge correlation spectrum.
To be specific, we discuss the system shown in Fig.~\ref{system},
namely a normal ring divided into two regions by tunnel barriers and
coupled capacitively to an external electrical circuit.  Each of the
two regions can be described by a single potential, and the system
considered permits thus a simple discussion of its polarizability.

We turn our attention to the specific case where the external
im\-pe\-dance $Z_{\mbox{\scriptsize\it ext}}(\omega)$ shows resistive
behavior at low frequencies.  We consider the model discussed in
\cite{cedraschi:prl}, namely an impedance modeled by a transmission
line.  We investigate the persistent current, the charge on the dot
and the flux induced capacitance which is the response of the charge
to a variation in the magnetic flux through the ring.  We identify two
different classes of ground states as a function of the coupling
between the external circuit and the ring, namely a ``Kondo'' type
ground state at weak coupling and a symmetry broken ground state at
strong coupling.  In order to understand better the different regimes,
we discuss time dependent quantities, in particular the power spectrum
of the charge fluctuations on the dot.

The case of a resistive external circuit is of particular importance
since the ring-dot structure coupled to a resistance exhibits
dephasing at finite temperatures \cite{cedraschi:langevin}.  In view
of the recent controversy concerning the question of whether there is
dephasing at zero temperature \cite{mohanty:decoh,golubev:decoh,%
golubev:decoh:reply} or not \cite{aleiner:decoh:comment,%
aleiner:decoh}, it is interesting to ask what we can learn about the
ground state of our simple system if it is coupled to a resistive
external circuit.  The case where the dot and the arm of the ring are
in resonance has already been treated in \cite{cedraschi:prl}.  Below,
we fill in the details left out in the discussion presented there and
extend it to other thermodynamic quantities as well as to the noise
spectrum of the charge on the dot.  The discussion of the latter is of
a particular interest since it exhibits decay of the correlations with
time {\em in the ground state}.  Here we emphasize an approach which
treats the entire system in a Hamiltonian fashion.  A more intuitive
description of this system based on a Langevin equation approach has
been given elsewhere \cite{cedraschi:langevin}.

\section{Resistive external circuit}
\label{extcircuit:resistor}
\begin{figure}
\centerline{\epsfysize=2cm\epsfbox{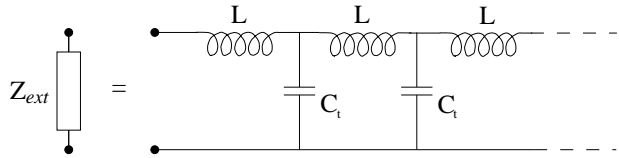}}
\caption{\label{trans}Transmission line with impedance
$Z_{ext}(\omega) = i\omega L/2 + \sqrt{L/C_t - \omega^2 L^2/4}$.}
\end{figure}

We model the finite impedance $Z_{\mbox{\scriptsize\it ext}}$ in a
Hamiltonian fashion, with the help of a transmission line with
capacitance $C_t$ and impedance $L$, see Fig.~\ref{trans} and
Appendix~A.  The ohmic resistance generated by the transmission line
is $R = Z_{\mbox{\scriptsize\it ext}}(\omega=0) = (L/C_t)^{1/2}$.  We
denote the charge operators on the capacitors between the inductances
by $\hat{Q}_n$ and introduce the conjugate phase operators
$\hat{\phi}_n$ fulfilling the commutation relations
\begin{equation}
[\hat{\phi}_m, \hat{Q}_n] = ie\delta_{mn}.
\end{equation}
It is via these charges and phases that the external circuit couples
to the ring.  The charge on the dot is $\hat{Q}_d$.  We define the
parallel internal capacitance $C_i \equiv C_L+C_R$, the external
series capacitance $C_e^{-1} \equiv C_1^{-1}+C_2^{-1}$ and the total
capacitances $C \equiv C_i+C_e$ and $C_0^{-1} \equiv C_i^{-1} +
C_e^{-1}$.  Then the Hamiltonian including all electrical interactions
in the circuit reads, see Appendix~A,
\begin{equation}
\label{extcircuit:HC}
\hat{H}_C = {\hat{Q}_d^2 \over 2C_i}
+ {\hat{Q}_d \hat{Q}_0 \over C_i}
+ {\hat{Q}_0^2 \over 2C_0}
+ \hat{H}_{HO},
\end{equation}
where $\hat{H}_{HO}$ is the Hamiltonian of a bath of harmonic
oscillators describing the transmission line,
\begin{equation}
\label{extcircuit:HHO}
\hat{H}_{HO} = \sum_{n=1}^\infty \left\{
{\hat{Q}_n^2 \over 2C_t} +
\left( {\hbar \over e} \right)^2
{(\hat{\phi}_n - \hat{\phi}_{n-1})^2 \over 2L}
\right\} .
\end{equation}
The spectrum of $\hat{H}_{HO}$ is discussed in some detail in
Appendix~A.  Here, we merely state that it is dense.  Similar models
involving a two-level system coupled to a bath of harmonic oscillators
with a discrete set of fundamental frequencies are investigated in the
context of stimulated photon emission.  A known example is the
Jaynes-Cummings model \cite{jaynes:model}.  These models exhibit
recurrence effects such as wave function revival
\cite{eberly:revival,pfeifer:decay}.  Because our model has a dense
spectrum of fundamental frequencies of the harmonic oscillators, there
are no recurrence phenomena \cite{leggett:review}.

\subsection{The total Hamiltonian}
\label{extcircuit:totalH}
At low enough temperatures, we can describe the ring-dot system in
terms of just two charge states, namely the state with $N$ electrons
and the state with $N+1$ electrons in the dot.  This leads to a two
level system \cite{buettiker:ringdot,buettiker:ringdot:prl} which we
discuss briefly in Appendix~B, and which is described by the
Hamiltonian $\hat{S} \hat{H}_e^{\mbox{\scriptsize\it eff}}
\hat{S}^{-1}$.  The total Hamiltonian reads
\begin{equation}
\label{extcircuit:CL1}
\hat{H}^{\mbox{\scriptsize\it eff}}
\equiv
\hat{S} \hat{H}_e^{\mbox{\scriptsize\it eff}} \hat{S}^{-1}
+ \hat{H}_C
= { \hbar \varepsilon \over 2 } \sigma_z
- { \hbar \Delta_0 \over 2} \sigma_x
+ { e \over 2C_i } \sigma_z \hat{Q}_0
+ \hat{H}_{HO}.
\end{equation}
Let us briefly discuss the different terms of
Eq.~(\ref{extcircuit:CL1}).  The Pauli matrix $\sigma_z$ is related to
the charge on the dot by $\hat{Q}_d = (e/2)\sigma_z + e(N-N_+ +1/2)$,
where $e N_+$ is an effective background charge.  The detuning
$\hbar\varepsilon$ is the energy difference that an electron must
overcome if it wants to tunnel from the arm to the dot.  It contains
the one-particle energies $\epsilon_{d(N+1)}$ and $\epsilon_{aM}$, and
a term stemming from the Coulomb interactions,
Eq.~(\ref{extcircuit:HC}).  It reads
\begin{equation}
\hbar \varepsilon = \epsilon_{d(N+1)} - \epsilon_{aM}
+ { e^2 \over C } \left( N - N_+ + { 1 \over 2 } \right).
\end{equation}
The tunneling across the barriers between the dot and the arm is
mediated by $\hbar\Delta_0 \, \sigma_x/2$, where the tunneling
amplitude $\hbar\Delta_0$ depends on the flux $\Phi$ and is given in
Eq.~(\ref{ta}).  The term $(e/2C_i)\sigma_z\hat{Q}_0$ is responsible
for the capacitive coupling between the ring and the external circuit,
whose Hamiltonian $\hat{H}_{HO}$ is given in
Eq.~(\ref{extcircuit:HHO}).

The Hamiltonian $\hat{H}^{\mbox{\scriptsize\it eff}}$ given in
Eq.~(\ref{extcircuit:CL1}) is the Caldeira-Leggett model
\cite{caldeira:diss,leggett:review}.  To make the connection to this
model more explicit, we introduce the cut-off frequency $\omega_c =
(LC_t)^{-1/2}$ of the transmission line and the parameter $\alpha$
describing the strength of the coupling to the bath
\begin{equation}
\label{extcircuit:alpha}
\alpha \equiv {R \over R_K} \left( {C_0 \over C_i} \right)^2,
\end{equation}
where $R_K \equiv h/e^2$ denotes the quantum of resistance, and
$R=(L/C_t)^{1/2}$ is the zero-frequency impedance of the external
circuit.  The low energy behavior of a system described by the
Hamiltonian of Eq.~(\ref{extcircuit:CL1}) is determined by the
parameters $\alpha$, $\omega_c$, $\Delta_0$ and $\varepsilon$.  The
limit of a ring which is not coupled to an environment is recovered by
setting $\alpha=0$.  This limit corresponds to vanishing external
capacitances, $C_1=C_2=0$, implying that $C_0=0$.  Before discussing
in more detail the different regimes associated with different values
of those parameters, let us rewrite the coupling between the two level
system and the bath in a form that is more suitable for our purposes.

We introduce the unitary transformation
\begin{equation}
\hat{U} = \exp \left(
-{i \over 2} {C_0 \over C_i} \hat{\phi}_0 \sigma_z
\right).
\end{equation}
With this transformation, we define an equivalent Hamiltonian
\begin{equation}
\hat{H} \equiv
\hat{U} \hat{H}^{\mbox{\it\scriptsize eff}} \hat{U}{}^{-1}.
\end{equation}
A simple calculation yields $\hat{H} = \hat{H}_{TLS} + \hat{H}_{HO} +
\hat{H}_I$, where $\hat{H}_{HO}$ has already been defined in
Eq.~(\ref{extcircuit:HHO}), and
\begin{eqnarray}
\hat{H}_{TLS} &=& { \hbar \varepsilon \over 2 } \sigma_z, \\
\label{extcircuit:HI}
\hat{H}_I &=& - { \hbar \Delta_0 \over 2 }
\left[
\sigma_+ \exp\left( i{C_0 \over C_i} \hat{\phi}_0 \right)
+ \sigma_- \exp\left( -i{C_0 \over C_i} \hat{\phi}_0 \right)
\right].
\end{eqnarray}
The coupling between the ring and the external circuit is thus
mediated by the operator $\exp[-i(C_0/C_i)\hat{\phi}_0]$ which
increases the charge on the capacitor $C_1$ (see Fig.~\ref{system}) by
$e\, C_0/C_i$.  We emphasize that the charge on the capacitor $C_1$ is
in general not shifted by an integer multiple of the electron charge
when an electron tunnels between the dot and the arm of the ring.  For
a more thorough discussion, see Appendix~C.  We investigate the
correlator
\begin{equation}
\label{extcircuit:Ptdef}
P(t) \equiv \left\langle
\exp\left( i{C_0 \over C_i} \hat{\phi}_0(t) \right)
\exp\left( -i{C_0 \over C_i} \hat{\phi}_0(0) \right)
\right\rangle_0,
\end{equation}
in particular its long time behavior.  Here, $\hat{\phi}_0(t) \equiv
e^{i\hat{H} t/\hbar } \hat{\phi}_0 e^{-i\hat{H} t/\hbar}$ is the time
dependent phase operator, and $\langle \ldots \rangle_0$ denotes the
expectation value taken with respect to the ground state for $\Delta_0
=0$, see also Appendix~D.  It is shown in there that the correlator
$P(t)$ behaves asymptotically like
\begin{equation}
\label{extcircuit:Pt}
P(\tau) \sim \left( 1 + i\omega_c |t| \right)^{2\alpha},
\quad (t \to \infty).
\end{equation}
Eq.~(\ref{extcircuit:Pt}) is of central importance.  It allows to show
that at $\alpha = 1/2$ the problem can be mapped onto the exactly
solvable Toulouse limit, Sec.~\ref{extcircuit:Toulouse}.  Moreover, it
allows for a discussion of the thermodynamic properties of the problem
using perturbation theory, see Sec.~\ref{extcircuit:pert}.
Furthermore, the correlator $P(t)$ is closely related to the function
$P(E)$ encountered in the discussion of tunneling through a tunnel
barrier shunted by an impedance, see \cite{ingold:tunnel} and
Appendix~C.  The discussion of $P(E)$ is generally referred to as
``$P(E)$-theory''.

\section{Ground state properties}
\label{extcircuit:gs}
In this section, we discuss ground state properties of the ring-dot
structure coupled to a resistive external circuit.  We are mainly
interested in the persistent current.  However, we also discuss two
quantities related to the polarizability of the ring-dot structure,
namely the charge on the dot, $\hat{Q}_d$ and the {\em flux induced
capacitance\/} $C_\Phi$ \cite{buettiker:ringdot,%
buettiker:ringdot:prl} which is the response of the charge on the dot
to a variation in the magnetic flux $\Phi$, much in the same way as
the electrochemical capacitance (see, for instance
\cite{buettiker:admittance}) is the response of the charge to a
variation in the applied bias.

\subsection{Equivalent models}
\begin{figure}
\centerline{\epsfysize5.5cm\epsfbox{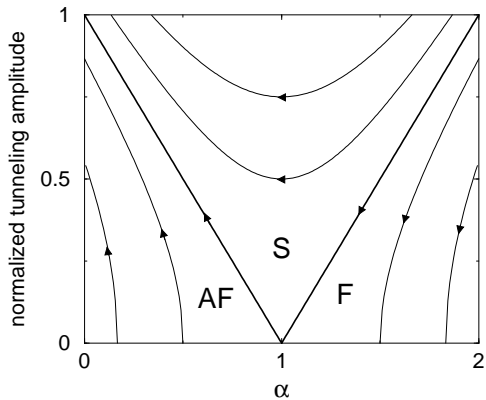}}
\caption{\label{extcircuit:sep} Scaling trajectories in the half plane
spanned by $\alpha$ and the normalized tunneling amplitude $\Delta_0 /
\omega_c$.  If the cut-off is lowered from $\omega_c $ to $\omega_c -
d\omega_c$, scaling takes place along the lines in the direction of
the arrows.  The bold lines correspond to the separatrix equation $| 1
- \alpha | = \Delta_0 / \omega_c$, dividing the plane up into the
strong tunneling (S), the ferromagnetic (F) and the anti-ferromagnetic
(AF) Kondo region.  This diagram has been investigated by
Anderson, Yuval, and Hamann \protect{\cite{anderson:kondo}}.}
\end{figure}

The anisotropic Kondo model, see {\em e.g.}\ \cite{hewson:kondo}, the
resonant level model \cite{schlottmann:reslev} and the inverse square
one-dimensional Ising model \cite{anderson:ising} are equivalent to
the Caldeira-Leggett model in their low energy properties.  For
vanishing magnetic field, corresponding to our model being at
resonance $\varepsilon = 0$, those models are characterized by two
parameters, which may be identified as $\alpha$ and $\Delta_0 /
\omega_c$.  The half plane spanned by $\Delta_0 / \omega_c \ge 0$ and
$\alpha$ is divided up into three regions by the separatrix equation
$|1-\alpha| = \Delta_0 / \omega_c$, where the parameters flow to
different fixed points under the action of the renormalization group
\cite{anderson:kondo}, see also Fig.~\ref{extcircuit:sep}.  The regime
of strong tunneling $\Delta_0 / \omega_c > |1-\alpha|$ is beyond the
scope of this work, and we shall not consider it any further.  Of the
remaining two, the region $\alpha>1$, $\Delta_0 / \omega_c < \alpha -
1$ corresponds to the ferromagnetic Kondo model where $\Delta_0 /
\omega_c$ renormalizes to zero, whereas the region $\alpha<1$,
$\Delta_0 / \omega_c < 1 - \alpha$ corresponds to the
anti-ferromagnetic Kondo model that has been solved by Bethe ansatz
\cite{tsvelick:kondo,andrei:kondo}.  For large detuning $\varepsilon$,
both cases can be treated by a perturbative expansion in $\Delta_0$.
The details of this expansion are treated in Appendix~D.

\subsection{Perturbative results}
\label{extcircuit:pert}
\begin{figure}
\centerline{\epsfysize=5.5cm\epsfbox{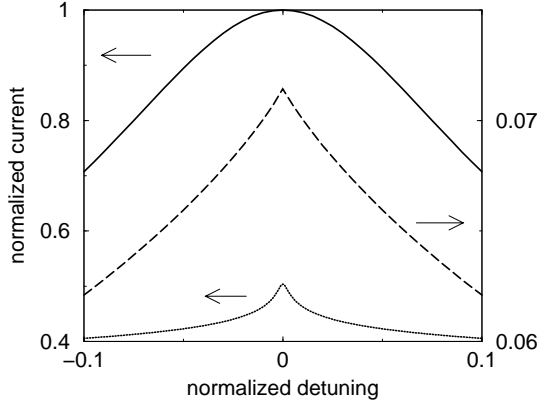}}
\caption{\label{extcircuit:pcbiased} The persistent current as a
function of the normalized detuning $\varepsilon / \omega_c$ for
$\alpha = 0$ (solid line), $\alpha = 1/2$ (dotted line) and $\alpha =
1.2$ (dashed line).  Units are chosen such that the persistent current
for $\alpha = 0$ is unity at resonance $\varepsilon = 0$.  Note the
different scale for the persistent current at $\alpha = 1.2$.  The
parameters are $\omega_c = 10\Delta_0$.}
\end{figure}

\begin{figure}
\centerline{\epsfysize5.5cm\epsfbox{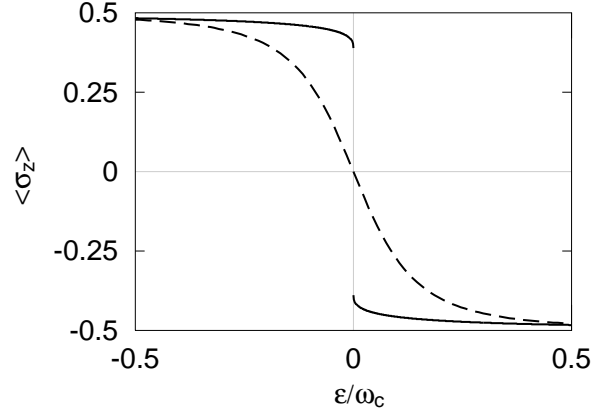}}
\caption{\label{extcircuit:jump}The position $\langle \sigma_z
\rangle$ of the electron for $\alpha=1.2$ (solid line) showing
symmetry breaking, compared to the typical behavior of $\langle
\sigma_z \rangle$ at $\alpha=0$ (dashed line).  The tunneling
amplitude $\Delta_0 / \omega_c=0.5$ is chosen very large in order to
make the bending near $\varepsilon=0$ visible in the case of symmetry
breaking.}
\end{figure}

At zero temperature and to second order in $\Delta_0$, the free energy
is $F_0 + \delta F$ with (see Appendix~D) $F_0 = - \hbar |\varepsilon|
/ 2$, and we find
\begin{equation}
\label{extcircuit:perturbativeF}
\delta F
= - { \hbar \omega_c \over 4 } \left( { \Delta_0 \over
\omega_c } \right)^2 e^{ | \varepsilon |/ \omega_c } \left| {
\varepsilon \over \omega_c } \right|^{ 2\alpha - 1 } \Gamma \left( 1 -
2\alpha, { | \varepsilon | \over \omega_c } \right) ,
\end{equation}
where $\Gamma(\zeta, x)$ is the incomplete gamma function.  If the
detuning $\varepsilon$ is large in magnitude, that is, far away from
resonance, Eq.~(\ref{extcircuit:perturbativeF}) is valid at any
$\alpha$ up to second order in $\Delta_0 / \omega_c$.  In the regime
$\alpha - 1 \gg \Delta_0 / \omega_c$ (the ferromagnetic regime of the
Kondo model, see Fig~\ref{extcircuit:sep}),
Eq.~(\ref{extcircuit:perturbativeF}) is valid even for arbitrary
detuning $\varepsilon$.  From Eq.~(\ref{extcircuit:perturbativeF}), we
easily obtain the persistent current in perturbation theory
\begin{equation}
\label{extcircuit:perturbativeI}
I = -c {\partial\delta F \over \partial\Phi}
= - {\hbar c \over 2} {\Delta_0 \over \omega_c}
{\partial\Delta_0 \over \partial\Phi}
e^{|\varepsilon|/\omega_c}
\left| {\varepsilon \over \omega_c} \right|^{2\alpha-1}
\Gamma \left( 1 - 2\alpha, {|\varepsilon| \over \omega_c} \right).
\end{equation}
Note that therefore for $\alpha > 1$, where
Eq.~(\ref{extcircuit:perturbativeI}) is valid for arbitrary detuning,
the persistent current has a cusp at resonance, see
Fig.~\ref{extcircuit:pcbiased}.  It is well known
\cite{chakravarty:sb,bray:sb} that the Caldeira-Leggett model exhibits
symmetry breaking for $\alpha > 1$, namely $\langle \sigma_z \rangle =
\partial F / \partial ( \hbar \varepsilon )$ exhibits a finite jump as
$\varepsilon$ crosses zero, see Fig.~\ref{extcircuit:jump}.  By virtue
of Eq.~(\ref{extcircuit:Qd}), $\langle \sigma_z \rangle$ is related to
the average charge $\langle \hat{Q}_d \rangle$ on the dot.  Let us
briefly illustrate this relation.  For $\varepsilon \to -\infty$,
$\langle \sigma_z \rangle$ goes to $+1$, and the charge on the dot
approaches therefore an integer multiple of the electron charge
$e(N+1-N_+)$.  As $\varepsilon$ increases and goes towards zero,
$\sigma_z$ approaches $0$ as well (unless there is symmetry breaking,
in which case it approaches a finite positive value).  At $\langle
\sigma_z \rangle = 0$, the charge on the dot is a half-integer
multiple of the electron charge, $\langle \hat{Q}_d \rangle =
e(N+1/2-N_+)$, and for $\varepsilon \to +\infty$, the charge on the
dot approaches again an integer multiple of $e$.  In the following, we
discuss for convenience the average $\langle \sigma_z \rangle$,
bearing in mind that the physically more meaningful quantity is the
charge on the dot $\langle \hat{Q}_d \rangle$.

The average $\langle \sigma_z \rangle$ is an odd function of
$\varepsilon$, and for $\varepsilon > 0$ it reads in perturbation
theory
\begin{eqnarray}
\left\langle \sigma_z \right\rangle
&=& - { 1 \over 2 }
+ { 1 \over 4 } \left( { \Delta_0 \over \omega_c } \right)^2
e^{ \varepsilon / \omega_c } \Bigg\{
\sum_{ n=0 }^\infty { (-1)^n \over n! }
{ ( \varepsilon / \omega_c )^n \over ( 1 - 2\alpha + n )
( 2 - 2\alpha +n ) } \nonumber\\
&-& \Gamma( 1 - 2\alpha )
\left( { \varepsilon \over \omega_c } \right)^{ 2\alpha - 2 }
\left[ { \varepsilon \over \omega_c } + 2\alpha - 1 \right] \Bigg\} ,
\end{eqnarray}
where we have used the series expansion for the incomplete gamma
function, see Eq.~(\ref{pathint:gamma}).  In the case of symmetry
breaking, $\alpha>1$, the width of the jump at $\varepsilon = 0$ is
thus
\begin{equation}
\label{extcircuit:sigmajump}
\langle \sigma_z \rangle_{\varepsilon = 0-}
- \langle \sigma_z \rangle_{\varepsilon = 0+}
= 1 - \left( { \Delta_0 \over \omega_c } \right)^2
{ 1 \over 2 (2\alpha-1) (2\alpha-2) },
\end{equation}
see also Fig.~\ref{extcircuit:jump}.  The jump in $\langle \sigma_z
\rangle$ and the cusp in the persistent current have the same origin,
thus the cusp is an indicator of symmetry breaking.

The sensitivity of the charge $\langle \hat{Q}_d \rangle$ on the dot,
see also Eq.~(\ref{extcircuit:Qd}), to changes in the magnetic flux
$\Phi$ can be characterized in terms of the flux induced capacitance
\cite{buettiker:ringdot:prl,buettiker:ringdot}
\begin{equation}
C_\Phi \equiv {\partial\langle\hat{Q}_d\rangle\over\partial\Phi}
= e { \partial \langle \sigma_z \rangle \over \partial \Phi },
\end{equation}
It is an odd function of $\varepsilon$ as well, and for $\varepsilon >
0$, it reads
\begin{eqnarray}
C_\Phi &=& {e \over 2}
{\Delta_0 \over \omega_c}
{\partial (\Delta_0/\omega_c) \over \partial\Phi}
e^{\varepsilon/\omega_c}
\Bigg\{ \sum_{ n=0 }^\infty { (-1)^n \over n! }
{ ( \varepsilon / \omega_c )^n
\over ( 1 - 2\alpha +n ) ( 2 - 2\alpha +n ) }
\nonumber\\
&-& \Gamma( 1 - 2\alpha )
\left( { \varepsilon \over \omega_c } \right)^{ 2\alpha - 2 }
\left[ { \varepsilon \over \omega_c } + 2\alpha - 1 \right] \Bigg\}.
\end{eqnarray}
Like the average electron number, Eq.~(\ref{extcircuit:sigmajump}), it
exhibits a finite jump at resonance
\begin{equation}
C_\Phi( \varepsilon = 0+ ) - C_\Phi( \varepsilon = 0- )
= e {\Delta_0 \over \omega_c}
{\partial (\Delta_0/\omega_c) \over \partial\Phi}
{ 1 \over ( 2\alpha - 1 ) ( 2\alpha - 2 ) } ,
\end{equation}
for $\alpha - 1 \gg \Delta_0 / \omega_c$.  Thus the symmetry breaking
can be seen as well in the flux induced capacitance, see
Fig.~\ref{extcircuit:capacitance}.  Flux sensitive screening in rings
has recently been observed by Deblock {\em et al.}\
\cite{deblock:screening}.

\subsection{Bethe ansatz results}
For $0 < \alpha < 1$, we use the Bethe ansatz solution of the resonant
level model \cite{ponomarenko:reslev}.  The low energy properties of
the problem depend on three energy scales, namely the detuning
$\varepsilon$, a cut-off $D$ and the ``Kondo temperature''
\begin{equation}
T_K = \Delta_0 \left( { \Delta_0 \over  D } \right)^{ \alpha \over
1 - \alpha } ,
\end{equation}
which is the only scale that depends on the magnetic flux $\Phi$.  The
persistent current is calculated from the known expression
\cite{ponomarenko:reslev} for $\langle \sigma_z \rangle = \partial F /
\partial (\hbar \varepsilon)$
\begin{equation}
\label{extcircuit:sigmabethe}
\left\langle \sigma_z \right\rangle 
= { i \over 4 \pi^{3/2} }
\int_{-\infty}^\infty { dp \over p - i0 } e^{ ip ( \ln z + b ) }
{ \Gamma( 1 + ip ) \Gamma( 1/2 - i(1-\alpha)p )
\over \Gamma( 1 + i\alpha p ) } ,
\end{equation}
where $z = ( \varepsilon / T_K )^{2(1-\alpha)}$ and $b = \alpha \ln
\alpha + (1-\alpha) \ln (1-\alpha)$.  This Bethe ansatz solution is
valid for $1-\alpha > \Delta_0 / \omega_c$.  Now, we make use of the
fact that in terms of the above-mentioned energy scales, and at zero
temperature, the free energy reads $F = \hbar T_K f(\varepsilon/T_K,
T_K/D)$, where $f$ is a universal function.  The persistent current $I
= -c (\partial F/\partial\Phi)$ may thus be expressed in terms of
$\langle \sigma_z \rangle = \partial F / \partial (\hbar \varepsilon)$
\begin{equation}
\label{extcircuit:avI} 
I = -\hbar c {\partial T_K \over \partial\Phi} \left[
\int_0^y dx \, \langle \sigma_z \rangle (x)
- y \langle \sigma_z \rangle ( y ) \right] + I_0 .
\end{equation}
where $y \equiv \varepsilon / T_K$.  The integration constant $I_0$ is
just the persistent current at resonance $\varepsilon=0$ and is
determined below.  The integral in Eq.~(\ref{extcircuit:avI}) can be
performed giving
\begin{eqnarray}
\label{extcircuit:pc1}
I &=& {\hbar c \over \sqrt{\pi}}
\left( { \Delta_0 \over \omega_c } \right)^{ \alpha \over 1 - \alpha }
{\partial \Delta_0 \over \partial\Phi} \nonumber\\
&\times&
\Bigg\{ \sum_{n=1}^\infty { (-1)^n \over (n-1)! }
{ \Gamma\left( { 1 \over 2 } + n ( 1 - \alpha ) \right) e^{ -nb }
\over ( 1 - 2n ( 1 - \alpha ) ) \Gamma( 1 - n\alpha ) }
\left( { \varepsilon \over T_K } \right)^{ 1 - 2n ( 1 - \alpha ) }
\nonumber\\
&-& { \Gamma\left( 1 - { 1 \over 2 ( 1 - \alpha ) } \right)
e^{ - { b \over 2 ( 1 - \alpha ) } }
\over 2 ( 1 - \alpha )
\Gamma\left( 1 - { \alpha \over 2 ( 1 - \alpha ) } \right) }
\Bigg\} + I_0 ,
\end{eqnarray}
for $\ln z > -b$.  Near resonance, more precisely for $\ln z < -b$, it
reads
\begin{eqnarray}
\label{extcircuit:pc2}
I &=& {\hbar c \over \sqrt{\pi}} {1 \over 2(1-\alpha)}
\left( { \Delta_0 \over \omega_c } \right)^{ \alpha \over 1 - \alpha }
{\partial\Delta_0 \over \partial\Phi} \nonumber\\
&\times&
\sum_{n=1}^\infty { (-1)^n \over n! }
{ \Gamma\left( 1 + { n - 1/2 \over 1 - \alpha } \right)
e^{ { n - 1/2 \over 1 - \alpha } b } \over
\Gamma\left( 1 + ( n - 1/2 ) { \alpha \over 1 - \alpha } \right) }
\left( { \varepsilon \over T_K } \right)^{2n} + I_0 .
\end{eqnarray}
For $\alpha = 1/2$, the series in Eq.~(\ref{extcircuit:pc2}) can be
explicitly re-summed
\begin{equation}
\label{extcircuit:pc3}
I\left( \alpha = { 1 \over 2 } \right)
= {\hbar c \over 16} { \Delta_0 \over \omega_c }
{\partial\Delta_0 \over \partial\Phi}
\ln\left[ 1 + \left( { 16 \over \pi } { \omega_c \over \Delta_0 }
{ \varepsilon \over \Delta_0 } \right)^2 \right] + I_0 .
\end{equation}
This reflects the fact that the case $\alpha = 1/2$ corresponds to an
exactly solvable limit of the Kondo model, the Toulouse limit
\cite{toulouse:limit}, see also Sec.~\ref{extcircuit:Toulouse}.  It
remains to calculate the persistent current at resonance, $I_0$.  We
do so in the following paragraph.

We point out that $T_K$ sets the scale for the transition from
resonant to perturbative behavior, {\em i.e.}\ for $|\varepsilon| \gg
T_K$, the expressions for the persistent current from Bethe ansatz and
from the perturbation theory must coincide up to terms of order
$\Delta_0$.  This observation allows us to determine the cut-off $D$
in terms of $\omega_c$,
\begin{equation}
\label{cutoff}
\left( { D \over \omega_c } \right)^{2\alpha}
= {2 \Gamma( 3/2 - \alpha ) e^{-b}
\over \sqrt{\pi} ( 1 - 2\alpha ) \Gamma( 1 - 2\alpha ) 
\Gamma( 1 - \alpha ) } ,
\end{equation}
as well as the integration constant $I_0$
\begin{equation}
\label{extcircuit:I0}
I_0 = - {\hbar c \over 2} 
{ \Gamma( 1 - { 1 \over 2 ( 1 - \alpha ) } )
e^{ - { b \over 2 ( 1 - \alpha ) } }
\over \sqrt{\pi} ( 1 - \alpha )
\Gamma( 1 - { \alpha \over 2 ( 1 - \alpha ) } ) }
\left( { \Delta_0 \over D } \right)^{ \alpha \over 1 - \alpha }
{\partial\Delta_0 \over \partial\Phi}
+ {\hbar c \over 2} { 1 \over 1 - 2 \alpha }
{\Delta_0 \over \omega_c} {\partial\Delta_0 \over \partial\Phi}.
\end{equation}
For $\alpha < 1/2$, the first term in Eq.~(\ref{extcircuit:I0}),
behaving like $( \Delta_0 / D )^{\alpha/(1-\alpha)}$ is dominating.
For $\alpha > 1/2$ the second one dominates and behaves as $\Delta_0 /
D $.  The power law for $\alpha > 1/2$ is thus the same as one obtains
from perturbation theory at $\alpha > 1$, using $I = -c\, \partial F /
\partial\Phi$ with $F$ given by Eq.~(\ref{extcircuit:perturbativeF}).
The transition from the anti-ferromagnetic to the ferromagnetic Kondo
model as $\alpha$ crosses 1 is therefore not seen in the persistent
current at resonance, although it is visible, as stated above, as a
cusp at resonance when the persistent current is traced as function of
the de\-tuning~$\varepsilon$.

Leggett {\em et al.}\ \cite{leggett:review} investigate coherence
properties by discussing the decay of a state which is prepared such
that the electron is {\em e.g.}\ in the dot for negative times $t<0$
(by applying a large negative detuning $\varepsilon$, for example) and
released at $t=0$ (by setting $\varepsilon = 0$ for $t \ge 0$).  They
find a decay with superimposed oscillations for $\alpha < 1/2$ and a
decay without oscillations for $\alpha > 1/2$.  This behavior is
interpreted as decoherence setting in at $\alpha = 1/2$.  See also our
discussion of the charge correlator in Sec.~\ref{corr}.

The persistent current, however, remains differentiable in
$\varepsilon$ as $\alpha$ passes through $1/2$ and the ground state
retains some coherence even for large $\alpha$, see also
Sec.~\ref{corr}. As a function of detuning the cusp at resonance shows
up only at $\alpha > 1$.  The poles at $\alpha = 1/2$ in the two terms
in Eq.~(\ref{extcircuit:I0}) cancel each other.  Together they give
rise to a logarithmic persistent current
\begin{equation}
I_0\left( \alpha = { 1 \over 2 } \right)
\approx \hbar c {\partial\Delta_0 \over \partial\Phi}
{ \Delta_0 \over \omega_c } 
\left( \ln { \Delta_0 \over \omega_c } - 0.217 \ldots \right).
\end{equation}
The logarithmic term $\ln(\Delta_0/\omega_c)$ characterizes the
transition from power law to linear behavior.  The poles in the first
term in Eq.~(\ref{extcircuit:I0}) appearing at $\alpha=(2n+1)/(2n+2)$
for $n\ge 0$ are canceled by terms of higher order in $\Delta_0$.

\begin{figure}
\centerline{\epsfysize5.5cm\epsfbox{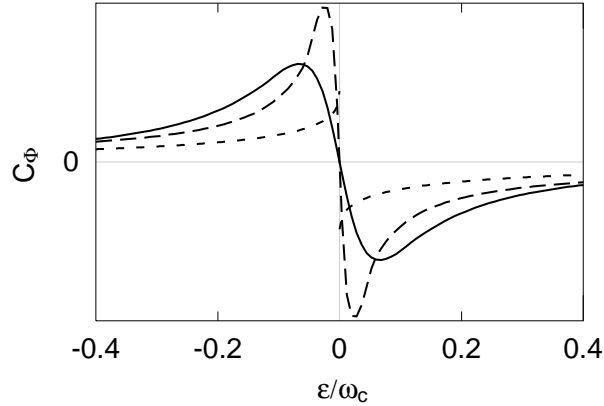}}
\caption{\label{extcircuit:capacitance}The flux induced capacitance as
a function of the detuning $\varepsilon/\omega_c$ and for a tunneling
amplitude $\Delta_0 / \omega_c = 0.2$.  The flux induced capacitance
is divided by $\partial\Delta_0 / \partial\Phi$ such that the graph is
independent of the flux $\Phi$, and given in arbitrary units.  The
solid line stands for $C_\Phi$ at $\alpha=0.3$, the long dashed line
is $C_\Phi$ for $\alpha=0.6$.  The short dashed line represents
$C_\Phi$ at $\alpha=1.2$ blown up by a factor $5$ and exhibits
symmetry breaking in $\varepsilon=0$.}
\end{figure}

The flux induced capacitance $C_\Phi$, see Sec.~\ref{extcircuit:pert},
can be obtained directly from the Bethe ansatz expression
Eq.~(\ref{extcircuit:sigmabethe}), giving for $\varepsilon > 0$
\begin{equation}
C_\Phi = {e \over \sqrt{\pi}}
{\partial\Delta_0/\partial\Phi \over \Delta_0}
\sum_{ n=1 }^\infty { (-1)^{ n-1 } \over (n-1)! }
{ \Gamma( { 1 \over 2 } + n ( 1 - \alpha ) ) e^{ -nb }
\over \Gamma( 1 - n\alpha ) }
\left( { \varepsilon \over T_K } \right)^{ -2n ( 1 - \alpha ) },\quad
\end{equation}
for $\ln z > -b$, where $z$ and $b$ have been defined below
Eq.~(\ref{extcircuit:sigmabethe}).  For $\ln z < -b$, it reads
\begin{equation}
C_\Phi = {e \over \sqrt{\pi} (1-\alpha)}
{\partial\Delta_0/\partial\Phi \over \Delta_0}
\sum_{ n=0 }^\infty { (-1)^n \over n! }
{ \Gamma( 1 + { n + 1/2 \over 1 - \alpha } )
e^{ { n + 1/2 \over 1 - \alpha } b }
\over \Gamma( 1 + ( n + 1/2 ) { \alpha \over 1 - \alpha } ) }
\left( { \varepsilon \over T_K } \right)^{ 2n + 1 } .
\end{equation}
There is no symmetry breaking for $\alpha < 1$, that is $C_\Phi = 0$
at resonance, see Fig.~\ref{extcircuit:capacitance}.

\section{Charge correlations}
\label{corr}
In this section, we consider charge-charge correlations on the dot at
unequal times.  These correlations may serve as an alternative measure
of the coherence properties of the ground state.  We define operator
$\Delta \hat{Q}_d(t)$ of the charge fluctuations on the dot by
\begin{equation}
\Delta \hat{Q}_d(t) \equiv \hat{Q}_d(t) - \langle \hat{Q}_d \rangle ,
\end{equation}
where the time evolution of the operator $\hat Q_d$ is given by
\begin{equation}
\hat{Q}_d(t) \equiv e^{ i\hat{H} t / \hbar }
\hat{Q}_d e^{ - i\hat{H} t / \hbar } .
\end{equation}
With Eq.~(\ref{extcircuit:Qd}) the symmetrized charge correlations may
be written in terms of the Pauli matrix $\sigma_z$
\begin{equation}
\label{corr:chcorr}
S_{QQ}(t) \equiv
{ 1 \over 2 } \left\langle \left\{ \Delta \hat{Q}_d(t),
\Delta\hat{Q}_d(0) \right\} \right\rangle
= { e^2 \over 8 } \left(
\left\langle \left\{ \sigma_z(t),
\sigma_z(0) \right\} \right\rangle
- 2 \left\langle \sigma_z \right\rangle^2 \right) ,\quad
\end{equation}
where the curly brackets $\{ \cdot, \cdot \}$ denote the
anti-commutator $\{ \hat{A}, \hat{B} \} = \hat{A} \hat{B} + \hat{B}
\hat{A}$.  In the following we will concentrate on two simple cases,
namely $\alpha = 0$ and $\alpha = 1/2$.

\subsection{Correlations at $\alpha = 0$}
The case $\alpha = 0$ is very easy to treat.  In this case the ring is
not coupled to the external circuit, and it suffices to calculate the
charge correlator, Eq.~(\ref{corr:chcorr}) for the ground state of the
$2 \times 2$-matrix $(\hbar\varepsilon/2) \sigma_z - (\hbar\Delta_0/2)
\sigma_x$, see Eq.~(\ref{extcircuit:CL1}), and we find
\begin{equation}
\label{corr:2t2}
S_{QQ}(t) = { e^2 \over 4 }
{ \Delta_0^2 \over \varepsilon^2 + \Delta_0^2 }
\cos \left( t \sqrt{ \varepsilon^2 + \Delta_0^2 } \right) .
\end{equation}
The correlations are periodically oscillating with time, and the
envelope of the oscillations is a constant.  The spectral density
$S_{QQ}(\omega)$, defined by
\begin{equation}
\label{corr:sqq}
S_{QQ}( \omega ) \equiv \int dt\,
e^{ i \omega t } S_{QQ}(t) ,
\end{equation}
shows $\delta$-peaks at $\omega = \pm \sqrt{ \varepsilon^2 +
\Delta_0^2 }$
\begin{equation}
S_{QQ}( \omega )
= \pi { e^2 \over 4 } { \Delta_0^2 \over \varepsilon^2 + \Delta_0^2 }
\left[ \delta( \omega - \sqrt{ \varepsilon^2 + \Delta_0^2 } )
+ \delta( \omega + \sqrt{ \varepsilon^2 + \Delta_0^2 } ) \right] .
\end{equation}
This also reflects the fact that there is no decay of the correlator
with time.

\subsection{Correlations at $\alpha = 1/2$}
\label{extcircuit:Toulouse}
\begin{figure}
\centerline{\epsfysize=5.5cm\epsfbox{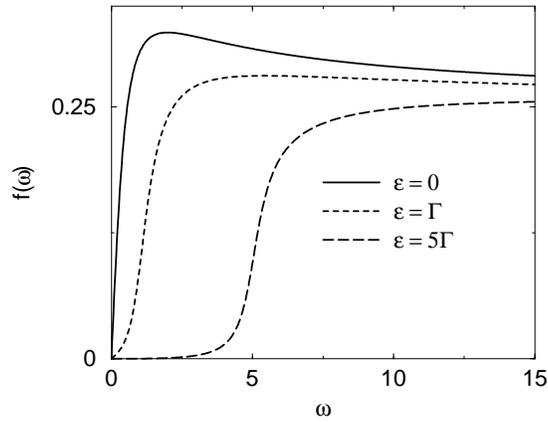}}
\caption{\label{corr:coeff}The function $f( \omega )$ describing the
deviation of the spectral density $S_{QQ}( \omega )$ at $\alpha = 1/2$
from the spectral density of an exponentially decaying correlator.
The frequency $\omega$ is measured in units of $\Gamma$.}
\end{figure}

\begin{figure}
\centerline{\epsfysize=5.5cm\epsfbox{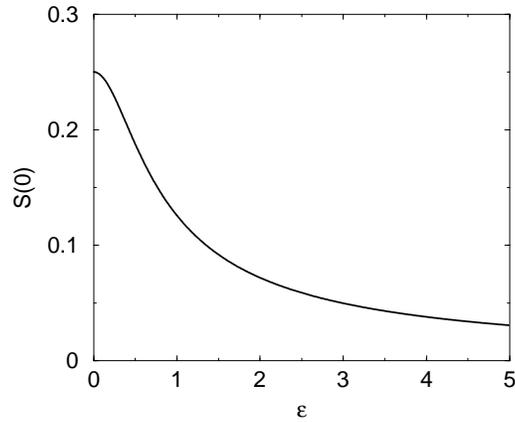}}
\caption{\label{corr:inst}Instantaneous charge correlations as a
function of $\varepsilon$, which is measured in units of $\Gamma$ at
$\alpha = 1/2$.  The function is symmetric under $\varepsilon \to
-\varepsilon$.  The correlations $S_{QQ}(t=0)$ (denoted by $S(0)$ in
the figure) are measured in units of the squared electron charge
$e^2$.}
\end{figure}

\begin{figure}
\centerline{\epsfysize=5.5cm\epsfbox{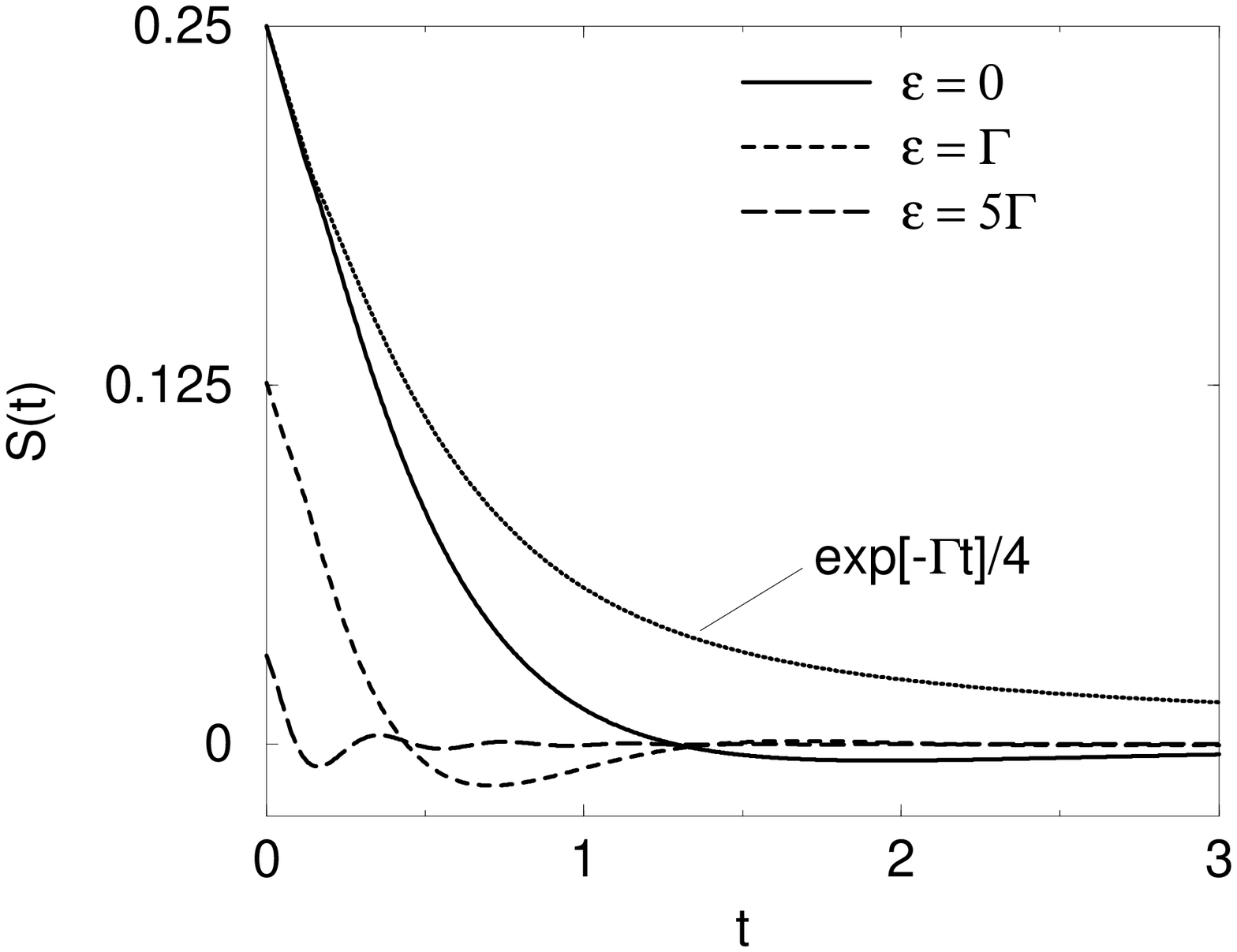}}
\caption{\label{corr:corr} Time dependence of the charge correlations
$S_{QQ}(t)$ (denoted by $S(t)$ in the figure) for different values of
$\varepsilon$ in units of the squared electron charge $e^2$ at $\alpha
= 1/2$.  The line showing the exponential decay $1/4\, \exp(-\Gamma
t)$ is given for comparison.  For $\varepsilon \neq 0$ the correlator
exhibits damped oscillations.}
\end{figure}

In this section, we set $\hbar = 1$.  In the case $\alpha = 1/2$ the
Caldeira-Leggett model can be mapped on an exactly solvable limit of
the Kondo model, the Toulouse limit \cite{toulouse:limit}.  Its
Hamiltonian describes the hybridization of a continuum of spinless
electrons with a localized electronic level but without
electron-electron interactions.  This can also be seen from the
correlator $P(t)$, see Eq.~(\ref{extcircuit:Pt}), which for $\alpha =
1/2$ takes the form
\begin{equation}
P_{\alpha=1/2}(t) \sim {1 \over 1 + i\omega_c |t|},
\quad (t \to \infty).
\end{equation}
This expression corresponds to the sum over all energies of the
Green's functions of free electrons in a symmetric wide band with
linear dispersion and a density of states $dn/d\omega =
\omega_c^{-1}$.  We denote the spectrum of the electrons in the
continuum by $\omega_n$ and their creation and annihilation operators
by $\hat{b}_n^\dagger$ and $\hat{b}_n$, resp.  The energy of the
localized level is $\epsilon$ and its creation and annihilation
operators are $\hat{c}^\dagger$ and $\hat{c}$.  The strength of the
hybridization is $J_\pm$.  The Toulouse Hamiltonian then reads
\begin{equation}
\label{corr:HToul}
\hat{H}_{\mbox{\scriptsize Toul}}
= \int dn \, \omega_n \hat{b}_n^\dagger \hat{b}_n
+ \epsilon\, \hat{c}^\dagger \hat{c}
+ { J_\pm \over 2 } \int dn \,
\left( \hat{c}^\dagger \hat{b}_n + \hat{b}_n^\dagger \hat{c} \right) ,
\end{equation}
Then the Caldeira-Leggett model at $\alpha = 1/2$ is mapped onto the
Toulouse limit by identifying the parameters, see
\cite{leggett:review}
\begin{eqnarray}
\varepsilon \, &\leftrightarrow& \, \epsilon \\
\Delta_0 \, &\leftrightarrow& \, 2^{-1/2} J_\pm .
\end{eqnarray}
Moreover, we identify the Pauli matrix $\sigma_z$ in the
Caldeira-Leggett model with the operator for the occupation of the
localized level
\begin{equation}
\label{corr:sz}
{ 1 \over 2} \sigma_z \leftrightarrow \hat{c}^\dagger \hat{c}
- { 1 \over 2 } .
\end{equation}
The Green's functions for this Hamiltonian can be calculated in closed
form.  In order to calculate the charge correlator,
Eq.~(\ref{corr:chcorr}), we need the time ordered Green's function of
the localized electron level
\begin{equation}
G_d(t) = -i \left\langle {\rm T} \hat{c}(t)
\hat{c}^\dagger (0) \right\rangle
\equiv -i \left[ \theta(t) \left\langle \hat{c}(t)
\hat{c}^\dagger (0) \right\rangle
- \theta(-t) \left\langle \hat{c}^\dagger(0) \hat{c}(t)  \right\rangle
\right], \quad
\end{equation}
where $\theta(t)$ is the Heaviside step function.  In Fourier space,
it takes the form
\begin{equation}
G_d( \omega ) = \int_{ -\infty }^\infty { d\omega \over 2\pi }
e^{ -i \omega t } G_d(t)
= \left[ \omega - \epsilon + i0
- \Sigma_d( \omega ) \right]^{-1} ,
\end{equation}
where the self-energy $\Sigma_d( \omega )$ reads
\begin{equation}
\Sigma_d( \omega ) = -i { \pi \over 2 \omega_c }
\left( { J_\pm \over 2 } \right)^2 {\rm sign}\, \omega
\equiv - { i \Gamma \over 2} {\rm sign}\, \omega ,
\end{equation}
where ${\rm sign}\,\omega$ denotes the sign of $\omega$.  We shall see
below that $\Gamma$ is in fact the inverse decay time of the charge
correlations.  In terms of the parameters of the Caldeira-Leggett
model, it reads
\begin{equation}
\Gamma = { \pi \over 2 } { \Delta_0^2 \over \omega_c } .
\end{equation} 
Using Eqs.~(\ref{corr:chcorr}) and (\ref{corr:sz}), and with the help
of Wick's theorem, we find the charge correlations for $\alpha = 1/2$
to be
\begin{equation}
\label{corr:SQQ}
S_{QQ}(t) = e^2 {\rm Re}\, \left[ G_d(t) G_d(-t) \right] .
\end{equation}
A rather long but simple calculation yields the spectral density
\begin{equation}
S_{QQ}( \omega ) = e^2 { 2 \Gamma \over \omega^2 + \Gamma^2 }
f( \omega ) ,
\end{equation}
with
\begin{eqnarray}
f( \omega ) &=& { 1 \over 4\pi } \left[
\arctan \left( { | \omega | - \varepsilon \over \Gamma / 2 } \right)
+ \arctan \left( { | \omega | + \varepsilon \over \Gamma / 2 } \right)
\right] \nonumber \\
&-& { 1 \over 8\pi } { \Gamma \over \omega } \left[
\ln { \varepsilon^2 + ( \Gamma / 2 )^2 \over
( \omega - \varepsilon )^2 + ( \Gamma / 2 )^2 }
+ \ln { \varepsilon^2 + ( \Gamma / 2 )^2 \over
( \omega + \varepsilon )^2 + ( \Gamma / 2 )^2 }
\right].\quad
\end{eqnarray}
Let us discuss $S_{QQ}(t)$.  We have, see also Fig.~\ref{corr:coeff},
\begin{equation}
f( \omega ) \to { 1 \over 4 } , \quad
\mbox{for $\omega \gg \max \{ \varepsilon, \Gamma \}$} ,
\end{equation}
and the Fourier transform of $2\Gamma / ( \varepsilon^2 + \Gamma^2 )$
is simply $e^{ -\Gamma |t| }$.  Moreover, for $t \to 0$, the
correlation function $S_{QQ}(t)$ approaches the integral of the
spectral density
\begin{equation}
S_{QQ}(0) = \int_{ -\infty }^\infty { d\omega \over 2\pi }
S_{QQ}( \omega ) ,
\end{equation}
which, for symmetry reasons, must equal $1/4$ for $\varepsilon = 0$.
A sketch of $S_{QQ}(0)$ is given in Fig.~\ref{corr:inst}.  The
correlation function thus reads for short times
\begin{equation}
S_{QQ}(t) \sim S_{QQ}(0) \, e^{ -\Gamma |t| } , \quad
(|t| \ll \min \{ \varepsilon^{-1}, \Gamma^{-1} \}).
\end{equation}
For long times, we use the stationary phase expansion and get
\begin{equation}
S_{QQ}(t) \sim - { e^2 \over \pi }
{ ( \Gamma / 2 )^2
\over \left[ \varepsilon^2 + ( \Gamma / 2 )^2 \right]^2 } t^{-2} ,
\quad (t \to \infty).
\end{equation}
Thus the correlations decay exponentially with rate $\Gamma$ at small
times and algebraically for large times.  Moreover, since
$S_{QQ}( \omega = 0 ) = 0$, the integral of $S_{QQ}(t)$ over
all times must vanish, therefore $S_{QQ}(t)$ takes negative values at
intermediate times, see Fig.~\ref{corr:corr}.  For $\varepsilon \neq
0$ the correlations are still oscillatory in time.

We have discussed here only the cases $\alpha=0$ and
$\alpha=1/2$ which correspond to no coupling at all between the ring
and the electric circuit ($\alpha=0$), and a relatively strong
coupling ($\alpha=1/2$).  At $\alpha = 0$, the charge correlation spectrum 
reveals simply the coherent 
charge oscillations between the dot and the arm 
with a sharp frequency. At $\alpha=1/2$, charge transfer still occurs, 
but it is now entirely characterized by the (flux-sensitive) 
relaxation rate $\Gamma$ and the 
detuning. Instead of the periodic charge transfer which exists 
for $\alpha=0$, the charge transfer has a much more stochastic character. 
Together with the results for the persistent
current and the polarizability of the ring obtained in the previous
section, the charge correlation spectrum, discussed here, provides 
an additional measure of the coherence properties 
of the ground state. 

\section{Conclusions}
We have investigated a simple model of a mesoscopic normal ring
coupled to an external electrical circuit.  We have shown that the
zero-point fluctuations of the external circuit influence measurable
quantities like the persistent current $I$, the flux induced
capacitance $C_\Phi$, and the polarization of the ring.  The reduction
of the persistent current as well as the onset of symmetry breaking,
observed in the polarization and in the flux induced capacitance
indicate that the coherence of the wave function in the ring is
reduced by fluctuations in the external circuit.  While such a
suppression of quantum coherence is not too surprising at finite
temperatures, we have demonstrated in the present paper that it
persists even in the extreme quantum limit of zero temperature.

\appendix{A}
\label{app:trans}
\appendixtitle{The Hamiltonian of the transmission line} In order to
understand the dynamics of the bath of harmonic oscillators describing
the transmission line shown in Fig.~\ref{trans}, we have to discuss
its properties in some detail.  Its impedance is of the form
$Z_{\mbox{\scriptsize\it ext}}(\omega) = i\omega L/2 + (L/C_t -
\omega^2 L^2/4)^{1/2}$, see {\em e.g.}\ \cite{feynman:lecturesII}.
This appendix is devoted to the discussion of the Hamiltonian of the
transmission line and its eigenstates.  We denote the charges on the
capacitors between the inductances by $Q_n$ and the potentials by
$V_n$, $n = 0, 1, 2, \ldots$.  The charge $Q_0$ plays a special role,
see Fig.~\ref{system}, because it is the charge sitting on the
capacitor $C_1$; $V_0$ is the corresponding potential.  Similarly,
$Q_\infty$ and $V_\infty$ are the charge and the potential on the
capacitor $C_2$.  It is via these charges and potentials that the
external circuit couples to the ring.  Furthermore, we introduce the
phases $\phi_n$, which satisfy the equations
\begin{equation}
{d\phi_n \over dt} = {e \over \hbar} \left( V_n-V_\infty \right).
\end{equation}
The canonically conjugate variables to the charges $Q_n$ are not the
phases $\phi_n$ but rather the generalized fluxes $(\hbar/e)\phi_n$.
We choose the dimensionless quantities $\phi_n$ for later convenience.
In terms of these variables the Hamiltonian of the transmission line
reads
\begin{equation}
H_{HO} = \sum_{n=0}^\infty \left\{
{Q_n^2 \over 2C_t}
+ \left( {\hbar\over e} \right)^2
{(\phi_{n+1}-\phi_n)^2 \over 2L}
\right\},
\end{equation}
which is canonically quantized by replacing the charges and phases by
operators $\hat{Q}_n$ and $\hat{\phi}_n$ and by imposing the
commutation relations $[\hat{\phi}_m,\hat{Q}_n] = ie\,\delta_{mn}$,
giving Eq.~(\ref{extcircuit:HHO}).  We note that $Q_n$ plays the role
of a momentum, and $\phi_n$ plays the role of a position.  In
particular, the quadratic form involving the charge $Q_n$ is positive
definite.  Due to a theorem of linear algebra, any pair of quadratic
forms, one of which is positive definite, can be simultaneously
diagonalized.  Therefore, the Hamiltonian $\hat{H}_{HO}$ of
Eq.~(\ref{extcircuit:HHO}) can be diagonalized by a unitary
transformation mapping $\hat{Q}_n$ onto $\hat{Q}(x)$, and
$\hat{\phi}_n$ onto $\hat{\phi}(x)$.  The transformed operators
$\hat{Q}(x)$ and $\hat{\phi}(x)$ satisfy the commutation relations
$[\hat\phi(x), \hat{Q}(y)] = ie\,\delta(x-y)$.  In this new basis
$\hat{H}_{HO}$ reads
\begin{equation}
\label{HHOdiag}
\hat{H}_{HO} = \int_0^1 dx\, \left\{
{\hat{Q}^2(x) \over 2C_t}
+ \left( {\hbar\over e} \right)^2 {2\over L}
\sin^2\left( {\pi x \over 2} \right) \hat{\phi}^2(x)
\right\}.
\end{equation}
In terms of the operators $\hat{Q}(x)$, the charge on the dot
$\hat{Q}_0$ is given by
\begin{equation}
\hat{Q}_0 = \sqrt{2} \int_0^1 dx\,\cos\left({\pi x\over 2}\right)
\hat{Q}(x),
\end{equation}
and analogously, we have
\begin{equation}
\label{phi0}
\hat{\phi}_0 = \sqrt{2} \int_0^1 dx\,\cos\left({\pi x\over 2}\right)
\hat{\phi}(x).
\end{equation}
Eq.~(\ref{phi0}) will be useful in Appendix~D in order to calculate
the correlator $P(\tau)$.  The diagonalized Hamiltonian
Eq.~(\ref{HHOdiag}), being a sum over harmonic oscillators, can be
written in terms of the bosonic creation and annihilation operators
$\hat{a}_x^\dagger$ and $\hat{a}_x$ which are related to the charges
$\hat{Q}(x)$ and the phases $\hat{\phi}(x)$ by
\begin{eqnarray}
\hat{\phi}(x) &=& { e \over i \hbar \sqrt{ 2 \lambda_x } }
\left( \hat{a}_x - \hat{a}_x^\dagger \right) , \\
\hat{Q}(x) &=& \sqrt{ \lambda_x \over 2 }
\left( \hat{a}_x + \hat{a}_x^\dagger \right) ,
\end{eqnarray}
with $\lambda_x = \hbar (C_t/L)^{ 1/2 } \sin(\pi x/2)$, and which obey
the commutation relations $[\hat{a}_x, \hat{a}_y^\dagger] =
\delta(x-y)$.  In terms of the operators $\hat{a}_x$ and
$\hat{a}_x^\dagger$, the Hamiltonian of the transmission line,
Eq.~(\ref{HHOdiag}) reads
\begin{equation}
\label{HHOa}
\hat{H}_{HO} = \int_0^1 dx\,\hbar\omega_x
\left( \hat{a}_x^\dagger \hat{a}_x + {1\over 2} \right),
\end{equation}
the eigenfrequencies $\omega_x$ being
\begin{equation}
\label{HOspectrum}
\omega_x = \sqrt{1\over LC_t}\sin{\pi x\over 2}.
\end{equation}
We note that the spectrum Eq.~(\ref{HOspectrum}) is linear at low
frequencies with a density of states $dx/d\omega \equiv
(2/\pi)\omega_c^{-1}$, where $\omega_c = (LC_t)^{-1/2}$, and that it
is cut off at a frequency $\omega_c$.  We shall need Eqs.~(\ref{HHOa})
and (\ref{HOspectrum}) in Appendix~D where we write the partition
function of the Caldeira-Leggett model as a path integral.

\appendix{B}
\label{twolev}
\appendixtitle{Degrees of freedom in the ring} The Hamiltonian
$\hat{H}_C$ of the Coulomb interactions, Eq.~(\ref{extcircuit:HC}),
has to be augmented by a Hamiltonian describing the electronic degrees
of freedom inside the ring.  We are interested in the effect of the
long range nature of the Coulomb interactions, therefore we disregard
charge redistribution and correlation effects and assume the
potentials in the dot and the arm of the ring to be constant over the
extent of the respective subsystem.  As a consequence, the electronic
wave functions are independent of the number of electrons in the dot
and the arm.  Moreover, we consider spinless electrons here.  It has
been demonstrated in \cite{buettiker:ringdot,buettiker:ringdot:prl}
that the inclusion of spin does not change the calculations a great
deal although it somewhat changes the physics of the problem.  In
particular, the Kondo effect for an isolated ring sets in only for a
tunneling amplitude $\hbar\Delta_0$ that is comparable to or larger
than the mean level spacing in the dot or the arm, whereas in this
paper we assume that $\hbar\Delta_0$ is much smaller than the mean
level spacings.  For a discussion of the Kondo effect in normal metal
rings, see the articles of Eckle {\em et al.}\ \cite{eckle:kondoring}
and of Kang and Shin \cite{kang:kondoring}.  We denote the creation
operator for an electron of energy $\epsilon_{dj}$ in the dot by
$\hat{d}_j^\dagger$ and the creation operator for an electron of
energy $\epsilon_{ak}$ in the arm by $\hat{a}_k^\dagger$.  The
tunneling through the left barrier is described by a hopping amplitude
$t_{jk}^L$, the tunneling through the right barrier by $t_{jk}^R$.
The total tunneling amplitude includes the phase picked up by an
electron whose wave function encloses the magnetic flux $\Phi$ and
reads
\begin{equation}
\label{extcircuit:tjk}
t_{jk}(\Phi)
\equiv t_{jk}^L + t_{jk}^R
\exp\left( 2\pi i {\Phi \over \Phi_0} \right).
\end{equation}
where $\Phi_0=hc/e$ is the elementary flux quantum.  As the dependence
of the Hamiltonian on the magnetic field is shifted into the boundary
conditions by Eq.~(\ref{extcircuit:tjk}), the tunneling amplitudes
$t_{jk}^L$ and $t_{jk}^R$ can be taken to be real.  Their relative
sign depends on the number of nodes of the wavefunctions they couple.
It is positive if the sum of the number of the nodes of the
wavefunction in the dot with energy $\epsilon_{dj}$ and of the number
of nodes of the wavefunction in the arm with energy $\epsilon_{ak}$ is
even, and negative otherwise.  This makes up for the parity effect in
the persistent current.  Now, we can write a noninteracting
Hamiltonian for the electronic degrees of freedom
\begin{equation}
\hat{H}_e = \sum_j \epsilon_{dj} \hat{d}_j^\dagger \hat{d}_j
+\sum_k \epsilon_{ak} \hat{a}_k^\dagger \hat{a}_k
+\sum_{jk} \left[
t_{jk}(\Phi) \hat{d}_j^\dagger \hat{a}_k + h.c.
\right] .
\end{equation}
All the many-body interactions are taken into account by $\hat{H}_C$
of Eq.~(\ref{extcircuit:HC}).  The operator for the charge on the dot
becomes
\begin{equation}
\hat{Q}_d = e\sum_j \hat{d}_j^\dagger \hat{d}_j - eN_+ ,
\end{equation}
with $e N_+$ being an effective background charge on the dot.  Next,
we make a crucial approximation, which allows us to write $\hat{H}_e$
and $\hat{Q}_d$ as $2\times 2$-matrices.

We shall assume in the following that the tunneling amplitudes through
the left and the right barriers are much smaller in magnitude than the
level spacing in the dot and the level spacing in the arm.  We note
that the number of charge carriers in the ring is conserved.
Following Stafford and one of the authors
\cite{buettiker:ringdot:prl,buettiker:ringdot}, we consider
hybridization between the topmost occupied electron level in the arm
and the lowest unoccupied electron level in the dot, $\epsilon_{aM}$
and $\epsilon_{d(N+1)}$ only.  To simplify the notation, we denote the
tunneling amplitudes between the levels $\epsilon_{aM}$ and
$\epsilon_{d(N+1)}$ by $t_L$ for tunneling through the left barrier
and by $t_R$ for tunneling through the right one, and we assume $t_L$
and $t_R$ to be positive.  The Hamiltonian for the electronic degrees
of freedom can be written in terms of the Pauli matrices $\sigma_z$
and $\sigma_\pm = (\sigma_x \pm i\sigma_y)/2$
\begin{eqnarray}
\label{extcircuit:Heff}
\hat{H}_e^{ \mbox{\scriptsize\it eff} }
&=& { \epsilon_{d(N+1)} - \epsilon_{aM} \over 2 } \sigma_z
- \left[
 t_L \pm t_R \exp\left( 2\pi i {\Phi \over \Phi_0} \right)
\right] \sigma_+ \nonumber\\
&-& \left[
 t_L \pm t_R \exp\left( -2\pi i {\Phi \over \Phi_0} \right)
\right] \sigma_-
+ E_0,
\end{eqnarray}
where $E_0 = \sum_{ j=1 }^N \epsilon_{dj} + \sum_{ k=1 }^{ M-1 }
\epsilon_{ak} + ( \epsilon_{aM} + \epsilon_{d(N+1)} ) / 2$.  The terms
in $\sigma_\pm$ describe the tunneling between the states
$\epsilon_{d(N+1)}$ and $\epsilon_{aM}$.  The relative sign of $t_L$
and $t_R$ is positive if the number $N+M$ of electrons in the ring is
odd, and negative if it is even, as discussed in the previous
paragraph.  The unitary transformation
$\hat{H}_e^{\mbox{\scriptsize\it eff}} \mapsto \hat{S}
\hat{H}_e^{\mbox{\scriptsize\it eff}} \hat{S}^{-1}$ where $\hat{S} =
\exp( -i \sigma_z \vartheta / 2 )$ and $\tan \vartheta = - t_R \sin
(2\pi\Phi/\Phi_0) / ( t_L + t_R \cos(2 \pi \Phi/\Phi_0))$, leaves the
term in $\sigma_z$ unchanged and transforms the tunneling term as
follows
\begin{equation}
\label{extcircuit:paulirot}
\hat{S} \left\{ \left[
t_L \pm t_R e^{2\pi i\Phi/\Phi_0}
\right] \sigma_+
+ \left[
t_L \pm t_R e^{-2\pi i\Phi/\Phi_0}
\right] \sigma_- \right\} \hat{S}^{-1}
={ \hbar \Delta_0 \over 2 } \sigma_x,\quad\,
\end{equation}
with
\begin{equation}
\label{ta}
{\hbar \Delta_0 \over 2}
= \sqrt{t_L^2 + t_R^2
\pm 2 t_L t_R \cos\left( 2\pi {\Phi \over \Phi_0} \right) },
\end{equation}
the positive sign applying again for an odd number of electrons, the
negative one for an even number of electrons in the structure.  The
charge on the dot, $\hat{Q}_d$, reads in the two level approximation
\begin{equation}
\label{extcircuit:Qd}
\hat{Q}_d = { e \over 2 } \sigma_z
+ e\left( N - N_+ + { 1 \over 2 } \right) .
\end{equation}

\appendix{C}
\label{pe}
\appendixtitle{Relation to $P(E)$-theory}
In this section we discuss the relation of the results obtained above
to the ``$P(E)$-theory'' \cite{ingold:tunnel}.  Usually, the
$P(E)$-theory is employed in order to calculate the tunneling rate
through a barrier which is shunted by an impedance $Z_{shunt}(\omega)$
in the Golden Rule approximation.  The system discussed here is
somewhat special in the sense that the impedance $Z_{ext}(\omega)$
does not act as a shunt, but is coupled only capacitively to the
tunnel barrier, see Fig.~\ref{system}.  Therefore, the electrons in
the ring cannot flow through the impedance as is the case for a shunt
impedance.  We can see the difference to ordinary $P(E)$-theory more
clearly by considering the Hamiltonian $\hat{H}_I$,
Eq.~(\ref{extcircuit:HI}), responsible for the coupling between the
ring and the transmission line.  The operator
$\exp[-i(C_0/C_i)\hat{\phi}_0]$ appearing in Eq.~(\ref{extcircuit:HI})
is a charge shift operator.  It differs from the charge shift operator
that is usually encountered in the context of tunneling in the
$P(E)$-theory in that it does not change the charge on the capacitor
$C_1$ by an integer multiple of the electron charge but by a fraction
$(C_0/C_i) e$,
\begin{equation}
e^{-i(C_0/C_i)\hat{\phi}_0} \hat{Q}_0 e^{i(C_0/C_i)\hat{\phi}_0}
= \hat{Q}_0 - {C_0 \over C_i} e.
\end{equation}
The difference is due to the fact that the standard $P(E)$-theory does
not consider $\hat{\phi}_0$ but an operator $\hat{\phi}$ which is
canonically conjugate to the charge $\hat{Q}_d$ on the dot.  This is
no problem in an open system, where the conjugate operator
$\hat{\phi}$ indeed exists.  In our closed system with only two charge
states, however, there is no conjugate operator to $\hat{Q}_d$,
therefore we stick to the operator $\exp[i(C_0/C_i) \hat{\phi}_0]$.
The coupling to the external circuit modifies the tunneling such that
the persistent current is reduced, as shown in the present paper.  The
modification is described by the correlator
\begin{equation}
\label{pe:Kt}
K(t) \equiv \left( {C_0 \over C_i} \right)^2
\left\langle
\hat{\phi}_0(t) \hat{\phi}_0(0) - \hat{\phi}_0^2(0)
\right\rangle_0,
\end{equation}
where $\langle \ldots \rangle_0$ denotes an expectation value with
respect to the ground state for $\Delta_0=0$, see also Appendix~D.
The pre-factor $(C_0/C_i)^2$ is characteristic for our problem and is
due to the fact that we are considering the phase operator
$\hat{\phi}_0$ of the external circuit.  The correlator $K(t)$ is
related to the function $P(t)$ given in Eq.~(\ref{extcircuit:Ptdef})
by $P(t) = \exp[K(t)]$, and the Fourier transform $P(E)$ of $P(t)$ has
given the theory its name.  For the purpose of calculations it is more
convenient, however, to stay with the correlator $K(t)$.  It has been
mentioned above, see Eq.~(\ref{extcircuit:Pt}), that $K(t) \sim
-2\alpha \ln(1+i\omega_c |t|)$ in the long time limit, where
$\omega_c$ is the cut-off frequency, and $\alpha$ is identical to the
coupling parameter given in Eq.~(\ref{extcircuit:alpha}), which
determines the suppression of the persistent current,
Eq.~(\ref{extcircuit:I0}).  In $P(E)$-theory, a general rule for
calculating $K(t)$ at finite temperature is given,
\begin{equation}
K(t) = 2\left( {C_0 \over C_i} \right)^2
\int_0^\infty {d\omega \over \omega}
{{\rm Re}\, Z_{ext}(\omega) \over R_K}
\left( e^{-i\omega t} - 1 \right)
\coth \left( {\hbar\omega \over 2kT} \right).
\end{equation}
In the zero temperature limit, this formula yields the same long time
behavior as the one obtained above for a special realization of the
transmission line.

\appendix{D}
\label{pathint}
\appendixtitle{Thermodynamic properties of the Caldeira-Leggett model}
We want to write the Caldeira-Leggett Hamiltonian $\hat{H} =
\hat{H}_{TLS} + \hat{H}_{HO} + \hat{H}_I$ in a form that is suitable
for the path-integral formalism.  For the bath of harmonic
oscillators, described by $\hat{H}_{HO}$, this has been done in
Eq.~(\ref{HHOa}).  The Hamiltonians containing the Pauli matrices
$\sigma_z$ and $\sigma_\pm$ are translated into the operator formalism
by introducing the fermion operators ${\hat c}^\dagger$, $\hat c$,
creating and annihilating an electron the dot, and
$\hat{\overline{c}}{}^\dagger$ and $\hat{\overline{c}}$, creating and
annihilating an electron in the arm
\begin{eqnarray}
\label{pathint:HTLS}
\hat{H}'_{TLS} &=& { \hbar \varepsilon \over 2 }
\left( \hat{c}^\dagger \hat{c}
- \hat{\overline{c}}{}^\dagger \hat{\overline{c}} \right) , \\
\label{pathint:HI}
\hat{H}'_I &=& - { \hbar \Delta_0 \over 2 }
\left[
\hat{c}^\dagger \hat{\overline{c}}
\exp\left( i{C_0 \over C_i} \hat{\phi}_0 \right)
+ \hat{\overline{c}}{}^\dagger \hat{c}
\exp\left( -i{C_0 \over C_i} \hat{\phi}_0 \right)
\right].
\end{eqnarray}
For the phase operator $\hat{\phi}_0$, responsible for the coupling
between the ring and the external circuit, we have
\begin{equation}
i\hat{\phi}_0 = {e \over \sqrt{C_t}}
\int_0^1 dx\, {\cos(\pi x/2) \over \sqrt{\hbar\omega_x}}
\left( \hat{a}_x - \hat{a}_x^\dagger \right).
\end{equation}
Thus the Hamiltonians $\hat{H}'_{TLS}$, $\hat{H}'_I$ and
$\hat{H}_{HO}$ can all be expressed in terms of the boson operators
$\hat{a}_x$ and $\hat{a}_x^\dagger$, and in terms of the fermion
operators $\hat{c}$, $\hat{c}^\dagger$ and $\hat{\overline{c}}$,
$\hat{\overline{c}}^\dagger$.

In order to identify the Hamiltonians $\hat{H}$ and $\hat{H}' \equiv
\hat{H}'_{TLS} + \hat{H}_{HO} + \hat{H}'_I$ we have to exclude the
case where both the dot and the arm level are occupied and the case
where both are empty, in other words, we have to exclude the subspaces
where the operator $\hat{n} \equiv \hat{c}^\dagger \hat{c} +
\hat{\overline{c}}{}^\dagger \hat{\overline{c}}$ takes the eigenvalues
0 and 2, resp.  However, the Hamiltonian $\hat{H}'$ does not change
the number of electrons in the ring,
\begin{equation}
[ \hat{H}', \hat{n} ] = 0.
\end{equation}
Therefore, the partition function $Z' \equiv {\rm Tr} \, \exp( -\beta
\hat{H}' )$ is of the form $Z' \equiv Z_0 + Z_1 +Z_2$, where the
subscript denotes the eigenvalue of $\hat{n}$, with $Z_1$ being equal
to the partition function $Z \equiv {\rm Tr} \, \exp( -\beta \hat{H}
)$ of the Caldeira-Leggett model.  Moreover, on the subspaces with
$\hat{n}=0$ and $\hat{n}=2$ the two level system and the bath of
harmonic oscillators do not interact, and on the same subspaces
$\hat{H}_{TLS}' = 0$.  Thus, $Z_0 = Z_2 = Z_{HO}$ with $Z_{HO} = {\rm
Tr} \, \exp( -\beta \hat{H}_{HO} )$, whereas $Z_1 = Z_{HO} Z_{TLS}
Z_I$.  Here, $Z_{TLS}$ is the partition function of $\hat{H}_{TLS}'$
restricted to the subspace where $\hat{n}=1$, or equivalently the
partition function of $\hat{H}_{TLS}$, which reads $Z_{TLS} = 2\,{\rm
cosh} (\beta\hbar\varepsilon/2)$.  The partition function $Z'$ reads
thus
\begin{equation}
Z' = 2 Z_{HO} \left( 1
+ Z_I \, {\rm cosh} \, { \beta \hbar \varepsilon \over 2 } \right) ,
\end{equation}
and the partition function $Z$ of the Caldeira-Leggett model becomes
consequently
\begin{equation}
\label{pathint:Z}
Z = 2 Z_{HO} Z_I {\rm cosh} \, { \beta \hbar \varepsilon \over 2 } .
\end{equation}
Eq.~(\ref{pathint:Z}) gives us the relation between the
Caldeira-Leggett model and the Hamiltonian Eqs.~(\ref{HHOa}) and
(\ref{pathint:HTLS},~\ref{pathint:HI}).  We shall exploit this
relation to evaluate the partition function of the Caldeira-Leggett
model in the path integral formalism.

For this purpose, we determine the partition function $Z' = {\rm Tr}
\exp(-\beta \hat{H}')$ in the path integral formalism.  The time
dependent creation and annihilation operators of the harmonic
oscillators become complex variables
\begin{eqnarray}
\hat{a}_x(t)
\equiv e^{ i\hat{H}t/\hbar } \hat{a}_x e^{ -i\hat{H}t/\hbar }
&\mapsto& a_x(t) , \\
\hat{a}_x^\dagger(t)
\equiv e^{ i\hat{H}t/\hbar } \hat{a}_x^\dagger e^{ -i\hat{H}t/\hbar }
&\mapsto& a_x^*(t) ,
\end{eqnarray}
and the operators for the electronic degrees of freedom become
Grassmann variables
\begin{eqnarray}
\hat{c}(t) \mapsto c(t),
&\quad& \hat{\overline{c}}(t) \mapsto \overline{c}(t) , \\
\hat{c}{}^\dagger (t) \mapsto c^*(t) ,
&\quad& \hat{\overline{c}}{}^\dagger (t)
\mapsto \overline{c}{}^* (t) .
\end{eqnarray}
We write the imaginary time ($\tau = it$) action functional $S =
S_{TLS} + S_{HO} + S_I$ with
\begin{eqnarray}
S_{TLS} &=& \int_0^\beta d\tau \, \left[ c^*(\tau)
\left(
{ \partial \over \partial \tau } + { \varepsilon \over 2 }
\right) c(\tau)
+\overline{c}{}^*(\tau)
\left( { \partial \over \partial \tau } - { \varepsilon \over 2 }
\right) \overline{c}(\tau) \right], \\
S_{HO} &=& \int_0^\beta d\tau \,
\int_0^1 dx \, a_x^*(\tau)
\left( { \partial \over \partial \tau } + \omega_x \right) a_x(\tau), \\
S_I &=& { \Delta_0 \over 2 } \int_0^\beta d\tau \, \Big[
c^*(\tau) \overline{c}(\tau)
\exp\left( i {C_0 \over C_i} \phi_0(\tau) \right)
\nonumber\\
&+& \overline{c}{}^*(\tau) c(\tau)
\exp\left( -i{C_0 \over C_i} \phi_0(\tau) \right)
\Big].
\end{eqnarray}
With these definitions, the partition function can be written as a
path integral \cite{ramond:FT}
\begin{equation}
Z'=\int {\cal D}c \, {\cal D}c^* \,
{\cal D}\overline{c} \, {\cal D}\overline{c}{}^* \,
{\cal D}a \, {\cal D}a^* \, e^{ -S } ,
\end{equation}
where ${\cal D}a$ and ${\cal D}a^*$ stand for integration over all
complex variables $a_x( \tau )$, $a_x^*( \tau )$, and ${\cal D}c,
\ldots$ etc.\ stand for integration over the Grassmann variables $c(
\tau ), \ldots$ etc.  The expectation value of a function $A (c, c^*,
\overline{c}, \overline{c}{}^*, \{ a_x \}, \{ a_x^* \})$ for vanishing
tunneling amplitude is given by
\begin{eqnarray}
\label{pathint:expv}
\left\langle A \right\rangle_0 &\equiv& { 1 \over Z_0 }
\int {\cal D}c \, {\cal D}c^*
\, {\cal D}\overline{c} \, {\cal D}\overline{c}{}^* \,
{\cal D}a \, {\cal D}a^* \nonumber\\
&\times&
A(c, c^*, \overline{c}, \overline{c}{}^*, \{ a_x \}, \{ a_x^* \} )
e^{ - \left( S_{TLS} + S_{HO} \right) } ,
\end{eqnarray}
where $Z_0$ denotes the partition function for vanishing tunneling
amplitude
\begin{equation}
\label{pathint:Z0}
Z_0 \equiv \int {\cal D}c \, {\cal D}c^*
\, {\cal D}\overline{c} \, {\cal D}\overline{c}{}^* \,
{\cal D}a \, {\cal D}a^* \,
e^{ - \left( S_{TLS} + S_{HO} \right) } ,
\end{equation}
thus the partition function $Z_I$ reads
\begin{equation}
\label{pathint:ZI}
Z_I = \left\langle e^{ -S_I } \right\rangle_0.
\end{equation}
This expression can be written as a series expansion in even powers of
$\Delta_0$.  It has been shown in \cite{cedraschi:langevin} that the
partition function thus obtained is formally equivalent to the
partition function of the anisotropic Kondo model
\cite{anderson:kondo}.  Below, we calculate the free energy to second
order in $\Delta_0$ from Eq.~(\ref{pathint:ZI}).  This perturbative
result is valid for large detuning $\varepsilon$ only, if $\alpha<1$,
and even at resonance $\varepsilon=0$ in the case of symmetry
breaking, $\alpha>1$.

Before calculating the free energy, we introduce the Green's functions
needed.  The electronic degrees of freedom are described by the
standard Matsubara Green's functions $G(\tau-\tau') \equiv \langle
c(\tau) c^*(\tau') \rangle_0$ and $\overline{G}(\tau-\tau') \equiv
\langle \overline{c}(\tau) \overline{c}^*(\tau') \rangle_0$, which are
{\em anti-periodic\/} with period $\beta$, that is, $G(\tau+\beta) =
-G(\tau)$ and $\overline{G}(\tau+\beta) = -\overline{G}(\tau)$.  On
the domain $-\beta \le \tau \le \beta$ they read \cite{mahan:manypart}
\begin{eqnarray}
\label{pathint:fermicorr1}
G(\tau) &=& e^{-\varepsilon\tau/2}
\left[ \theta(\tau)
- {1 \over e^{\beta\varepsilon/2} + 1} \right], \\
\label{pathint:fermicorr2}
\overline{G}(\tau) &=& -e^{\varepsilon\tau/2}
\left[ \theta(-\tau)
- {1 \over e^{\beta\varepsilon/2} + 1} \right].
\end{eqnarray}
where $\theta(\tau)$ is the Heaviside step function.  The relevant
operator $\hat{\phi}_0$ of the bath, see Eq.~(\ref{pathint:HI}), has
its dynamics described by the correlator
\begin{equation}
\left\langle
e^{i(C_0/C_i)\phi_0(\tau)}
e^{-i(C_0/C_i)\phi_0(0)}
\right\rangle_0
\equiv e^{K(\tau)},
\end{equation}
where $K(\tau) = (C_0/C_i)^2 \langle \phi_0(\tau)\phi_0(0) -
\phi_0^2(0) \rangle_0$.  This correlator, $\exp[K(\tau)]$ is
equivalent to the correlator $P(t)$,
Eqs.~(\ref{extcircuit:Ptdef},~\ref{extcircuit:Pt}), evaluated at
imaginary times.  The correlator $K(\tau)$ is easily computed for the
transmission line, Eq.~(\ref{HHOa}).  On the domain $-\beta \le \tau
\le \beta$ the Green's functions of the bosons $a_x$ read
\cite{mahan:manypart}
\begin{equation}
\label{pathint:bosoncorr}
\left\langle a_x(\tau) a_y^*(0) \right\rangle_0
= \delta(x-y) e^{-\omega_x\tau} \left[ \theta(\tau)
+ {1 \over e^{\beta\omega_x} - 1} \right].
\end{equation}
Being of bosonic nature, they are periodic in $\tau$ with period
$\beta$.  At zero temperature, $\beta \to \infty$, we may drop the
second term in square brackets in Eq.~(\ref{pathint:bosoncorr}) and
obtain the correlator for $\phi_0(t)$
\begin{equation}
\label{pathint:corr}
\left\langle \phi_0(\tau) \phi_0(0) \right\rangle_0
= {e^2 \over C_t} \int_0^1 dx \,
{e^{-\omega_x|\tau|} \cos^2(\pi x/2)  \over \hbar \omega_x }.
\end{equation}
We linearize the spectrum of the harmonic oscillators
\begin{equation}
\hbar \omega_x \to {\pi\over 2} \hbar \omega_c x,
\end{equation}
see also the discussion of the spectrum of the harmonic oscillators
following Eq.~(\ref{HOspectrum}).  The cosine in the integral in
Eq.~(\ref{pathint:corr}) is replaced by 1, and the integral is cut off
at a frequency $\omega_c$.  We find
\begin{equation}
\left\langle \phi_0(\tau) \phi_0(0) \right\rangle_0
\sim {4R \over R_K}
\int_0^\infty { d\omega \over \omega } e^{ -\omega |\tau| }
e^{ -\omega / \omega_c } \quad ( \tau \to \infty ),
\end{equation}
where we have used the quantum of resistance $R_K=h/e^2$ and the
resistance of the transmission line $R=(L/C_t)^{1/2}$.  This integral
is infrared divergent, a disease remedied by subtracting the
correlation function $\langle \phi_0^2(0) \rangle$.  We obtain the
correlator $K(\tau)$
\begin{equation}
K(\tau) \sim -2\alpha \int_0^\infty {d\omega \over \omega}
\left[ 1 - e^{-\omega |\tau|} \right] e^{-\omega/\omega_c}
= -2\alpha \ln \left( 1 + \omega_c |\tau| \right) \;\;
(\tau \to \infty),
\end{equation}
thereby defining the dimensionless parameter $\alpha$ describing the
strength of the coupling between the two level system and the bath
\begin{equation}
\alpha = {R \over R_K} \left( {C_0 \over C_i} \right)^2.
\end{equation}
From $K(\tau)$, we can calculate thermodynamical properties from the
partition function.  We shall do so in the next paragraph.

From Eq.~(\ref{pathint:Z}), we can calculate the free energy by the
relation $ -\beta F = \ln Z$.  In fact, the partition function of the
bath, $Z_{HO}$, only gives rise to an additive constant which we shall
neglect below.  We therefore write $F = F_0 + \delta F$ with
\begin{equation}
F_0 = -k_B T \ln Z_{TLS} \to - { \hbar |\varepsilon| \over 2 }
\quad ( T \to 0 ),
\end{equation}
where $k_B$ is the Boltzmann constant, and $T$ is the temperature.
The correction to the free energy due to the presence of the bath is
\begin{equation}
\label{pathint:deltaF}
\delta F = - k_B T \ln Z_I.
\end{equation}
We shall use the cumulant expansion of the expression for $Z_I$,
Eq.~(\ref{pathint:ZI}), to calculate $\delta F$ to second order in
$\Delta_0$
\begin{equation}
\label{pathint:cumulant}
Z_I = \left\langle e^{ -S_I } \right\rangle_0
=\exp\left[ - \left\langle S_I \right\rangle_0
+ {1\over 2} \left( \left\langle S_I^2 \right\rangle_0
- \left\langle S_I \right\rangle_0^2 \right) - \ldots \right].
\end{equation}
As $\langle S_I \rangle_0 = 0$, we only need $\langle S_I^2
\rangle_0$.  We find
\begin{equation}
\label{pathint:SIsq}
\left\langle S_I^2 \right\rangle_0
= -\Delta_0^2
\int_0^\beta d\tau \int_0^\beta d\tau' \,
G(\tau-\tau') \overline{G}(\tau'-\tau) e^{K(\tau'-\tau)}.
\end{equation}
Now, we use Eqs.~(\ref{pathint:cumulant},~\ref{pathint:deltaF}), to
calculate the correction $\delta F$ to the free energy to second order
in $\Delta_0$.  We assume for convenience that $\varepsilon > 0$.  The
case $\varepsilon < 0$ follows immediately from the symmetry of the
free energy
\begin{equation}
\label{pathint:symmetry}
F(-\varepsilon) = F(\varepsilon) .
\end{equation}
For $\varepsilon > 0$, we insert the zero temperature limits of the
fermion Green's functions
Eqs.~(\ref{pathint:fermicorr1},~\ref{pathint:fermicorr2}) into
$\langle S_I^2 \rangle_0$, Eq.~(\ref{pathint:SIsq}), and obtain
\begin{equation}
\delta F =
-{1 \over 2} {\left\langle S_I^2 \right\rangle_0 \over \beta}
= -{\Delta_0^2 \over 2}
\left( 1 + {1 \over \beta} {d \over d\varepsilon} \right)
\int_0^\beta d\vartheta \,
e^{-\varepsilon \vartheta}
(1 + \omega_c \vartheta)^{-2\alpha},\quad
\end{equation}
and take the limit of zero temperature, $\beta \to \infty$.  Taking
into account the symmetry relation, Eq.~(\ref{pathint:symmetry}), this
gives
\begin{equation}
\delta F = - { \hbar \omega_c \over 4 }
\left( { \Delta_0 \over \omega_c } \right)^2
e^{ |\varepsilon| / \omega_c }
\left| { \varepsilon \over \omega_c } \right|^{ 2\alpha - 1 }
\Gamma \left(1 - 2\alpha, { |\varepsilon| \over \omega_c } \right),
\end{equation}
where $\Gamma( \zeta, x )$ denotes the incomplete gamma function.
Using the series expansion for $\Gamma(\zeta,x)$
\begin{equation}
\label{pathint:gamma}
\Gamma(\zeta, x) = \Gamma(\zeta) - \sum_{n=0}^\infty {(-1)^n x^{\zeta
+ n} \over n! (\zeta + n)},
\end{equation}
and the shorthand notation $x \equiv \varepsilon / \omega_c$ this
becomes
\begin{equation}
\delta F = - { \hbar \omega_c \over 4 }
\left( { \Delta_0 \over \omega_c } \right)^2 e^{ |x| }
\left[ |x|^{2\alpha - 1} \Gamma \left( 1 - 2\alpha, |x| \right)
- \sum_{n=0}^\infty { (-1)^n |x|^n \over n! (1 - 2\alpha + n) }
\right] .
\end{equation}
This expression converges at resonance, $\varepsilon = 0$, for any
$\alpha > 1/2$
\begin{equation}
\delta F(\varepsilon = 0) = - {\hbar \omega_c \over 4( 2\alpha - 1) }
\left( {\Delta_0 \over \omega_c} \right)^2 .
\end{equation}
This result coincides, for $\alpha > 1$, with the numerical value
obtained by integrating the renormalization flow equations for the
ferromagnetic an\-iso\-tro\-pic Kondo model \cite{anderson:kondo}.
The expectation value $\langle \sigma_z \rangle = \partial F /
\partial (\hbar \varepsilon)$, however, diverges for $\alpha = 1$ in
perturbation theory.  Moreover, in perturbation theory, the condition
$-1/2 \le \langle \sigma_z \rangle \le 1/2$ is fulfilled only for
$\alpha > 1$ and $\Delta_0 / \omega_c \ll \alpha - 1$, which
corresponds to the ferromagnetic regime of the anisotropic Kondo
model, see Fig.~\ref{extcircuit:sep}.  It is in this domain we believe
the perturbation theory to apply for all $\varepsilon$.

\begin{acknowledgment}
This work was supported in part by the Swiss National Science
Foundation.
\end{acknowledgment}

\end{article}

\end{document}